\documentclass[12pt]{article}
\usepackage[english]{babel}
\usepackage{amsmath}
\usepackage{amssymb}
\usepackage{bbm}
\usepackage{color}
\usepackage{graphicx}
\usepackage{epsfig}
\usepackage[titletoc,toc,title]{appendix}
\usepackage[all]{xy}
\usepackage{float}
\Roman{section}
\newcommand{\ket}[1]{\left|#1\right>}
\newcommand{\bra}[1]{\left<#1\right|}
\newcommand{\C}{\mathbb{C}}
\newcommand{\Z}{\mathbb{Z}}
\begin{document}

\begin{center}
\textsc{\Large{Violation of Bell inequalities for multipartite systems}}
\newline

\large{Yanmin Yang}\footnote{yangym929@gmail.com}\\
\emph{\normalsize{Department of Mathematics, Guangzhou University,
 Guangzhou, P.R.China}}\\
\large{Zhu-Jun Zheng}\footnote{Zhengzj@scut.edu.cn}\\
\emph{\normalsize{Department of Mathematics, South China University of Technology, Guangzhou, P.R.China}}\\

\end{center}

\begin{abstract}
  In recent papers, the theory of representations of finite groups has been proposed  to analyzing the violation of Bell inequalities. In this paper, we apply this method to more complicated cases. For two partite system, Alice and Bob  each make one of $d$ possible measurements,  each measurement has $n$ outcomes.
 The  Bell inequalities based on the choice of two orbits are derived.
The classical bound is only dependent on the number of measurements $d$, but the quantum bound is dependent both on $n$ and $d$.  Even so, when $d$ is large  enough, the quantum bound is only dependent on $d$.
The subset  of  probabilities for four parties based on the choice of six orbits under group action is derived and its violation is described. Restricting  the six orbits to three parties by forgetting the last party, and guaranteeing the classical bound invariant,  the Bell inequality based on the choice of four orbits is derived. Moreover, all the corresponding nonlocal games are  analyzed.
\end{abstract}


\section{Introduction}

The  Bell inequalities are compelling examples of  essential differences between quantum and classical physics \cite{bell}. They  are characterized by three parameters, the number of parties($N$), the number of measurement  settings($M$) and the number of outcomes for each measurement($K$) \cite{clauser,clauser1,kaz,WW,CGLMP,SLK,cabello}.
The  famous Bell inequality, the Clauser-Horne-Shimony-Holt(CHSH) inequality \cite{clauser}, is for the case
$N=M=K=2$. The usual form of the CHSH inequality is:
$$
S=E(QS)+E(RS)+E(RT)-E(QT)\leq 2,
$$
where $Q,R$ are  measurements sent to  Alice by a referee,  $S,T$ are  are  measurements sent to  Bob by  same referee, Alice and Bob perform their measurements simultaneously and then return their results $+1$ or $-1$ to the referee,  $E(\cdot)$ is the expectation value of the product of the outcomes of the experiment. But, in quantum mechanics, the upper bound $2\sqrt{2}$ of $S$ can be attained  which is  larger  than $2$,  and CHSH violation is therefore predicted by the theory of quantum mechanics.
For general $N$ and $M=K=2$, the  Bell inequalities were  structured by Werner and Wolf  \cite{WW}.
For $N=2$, $M=2$, and general $K$, the Bell  inequalities  were found by Collins, Gisin, Linden, et al.\ \cite{CGLMP}, then Son, Lee and Kim  generalized this situation to multipartite arbitrary dimensional systems with $M=2$ \cite{SLK}.

Recently, there appeared interesting papers \cite{ugur}, \cite{ugur1}, \cite{Bolonek} and \cite{BS}. In these papers the method of group representations theories  has been proposed as a tool to analyzing the quantum mechanical violation of Bell inequalities.

In this paper, we apply the group  theory to analyzing the  violation of Bell inequalities for more complicated cases.
In Sec. \ref{sec2}, the scenario for two parties is considered. Alice and Bob share some state $\ket\phi$, and Alice performs one of $d$ measurements sent by a referee on her part of the state,
Bob does similar operation.
Then Alice and Bob return their measurement results $v_A(s)$ and $v_B(t)$ to referee, $v_A(s)$ and $v_B(t)$ take values in set $\{0,1,\cdots, n-1 \}$, arbitrarily $n$ is  a natural number. Two  chosen   orbits under an group action approach a Bell inequality. In this section, we will see that the classical bound is independent of $n$, but the quantum bound is dependent both on $n$ and $d$.  More interesting conclusion is that when the number of measurements is large  enough, the quantum bound is only dependent on $d$.
In Sec. \ref{sec3}, the cases of  $N=4$, $M=2$ and $K=4$ was analyzed. The Bell inequality is constructed based on the choice of group orbits. In Sec. \ref{sec4}, restricting  the six orbits in Sec. \ref{sec3} to three parties by forgetting the last party, and guaranteeing the classical bound 
 invariant,  the Bell inequality based on the choice of four orbits is derived for three partite system.
For all scenarios, the corresponding nonlocal games are all analyzed.

\section{Two partite systems}\label{sec2}

Suppose we have two parties, Alice and Bob, and their joint states are elements of a tensor product Hilbert space   $\mathbb{C}^n\otimes \C^n$, with each $\C^n$ is spanned by the orthonormal basis $\{\ket0, \ket1, \cdots, \ket{n-1}  \}$.
 Each of them can measure one of $d$ observables, and for each observable the possible values for the result of the measurement are $0$, $1$, $\cdots$ or $n-1$.  Alice's observables are $a_{j}$,
 Bob's are $b_{j}$, $j=0,1,\cdots,d-1$.

 For the $n\times n$ translation operator $T$, there has a spectral decomposition
\begin{eqnarray}\label{Tn}
 T &=& | w_0 \rangle\langle w_0|+e^{-i2\pi /n} | w_1 \rangle\langle w_1|+e^{-i4\pi /n} | w_2 \rangle\langle w_2| \nonumber\\
    &&  + \cdots + e^{i4\pi /n} | w_{n-2} \rangle\langle w_{n-2}|+ e^{i2\pi /n} | w_{n-1} \rangle\langle w_{n-1}|,
\end{eqnarray}
where
\begin{equation}\label{w_j2}
|w_j \rangle=\frac{1}{\sqrt{n}}\sum_{k=0}^{n-1} e^{i2\pi jk /n}\ket j, \hspace{5mm} \langle w_j|w_k\rangle=\delta_{jk}
\end{equation}
for any $j,k=0,1,\cdots,n-1$.
Chose an operator  $U$ such that $U^d=T$. Thus $\{j\rightarrow U^j \mid j=0,1,\cdots, nd-1 \}$ is a representation of the cyclic group $\mathbb{Z}_{nd}$, the group of integers $\Z$  modulo $nd$. We have $d$ basis $\{\ket{v_j^{k}}=U^k\ket j | j=0,1,\cdots,n-1 \}$, for any $k=0,1,\cdots,d-1$,
corresponding to Alice's observables $a_k$ and Bob's observables  $b_k$. $a_k=\sum_{j=0}^{n-1}j|v_j^{k}\rangle\langle v_j^{k}|$, similarly for $b_k$, $k=0,1,\cdots,d-1$.

Under the group action $\alpha:\{U^j\otimes U^j|j=0,1,\cdots,nd-1\}\times \C^n\otimes \C^n\longrightarrow \C^n\otimes \C^n$, $\alpha(U^j\otimes U^j,\ket \psi)=U^j\otimes U^j\ket \psi$, we choose two states:
\begin{equation}\label{2orbits}
  \ket{0v_1^0}\hspace{5mm} {\text and} \hspace{5mm} \ket{0v_1^1},
\end{equation}
each orbit has $nd$ elements, and the two orbits are distinct with each other. The sum of probabilities corresponding to these states give some   Bell inequalities. From local realistic theory, they read
\begin{equation}
\small{
 \begin{split}
&\sum_{k=0}^{n-1}\sum_{j=0}^{d-1} P(a_j=k,b_j=k\oplus1)+\sum_{k=0}^{n-1}\sum_{j=0}^{d-2} P(a_j=k,b_{j+1}=k\oplus1)\\
& +\sum_{k=0}^{n-1} P(a_{d-1}=k,b_{0}=k\oplus2) \leq 2d-1,
\end{split}}
\end{equation}
where $\oplus$ means the addition modulo $n$, $P(a_j=k,b_j=k\oplus1)$ means the probability of the event when Alice take measure $a_j$ and obtain  value $k$, Bob take measure $b_j$ and obtain  value $k\oplus1$.
But in  quantum mechanics, the bound of the sum of these probabilities can attain a larger value.

In order to get the  quantum mechanics bound, we need to find a special state $\ket\phi$, such that the expectation value $\langle \phi |O | \phi\rangle$ is maximum, where
\begin{equation}\label{Belleq O2}
  O= \sum_{j=0}^{nd-1}(U\otimes U)^j(\ket{0v_1^0}\bra{0v_1^0}+\ket{0v_1^1}\bra{0v_1^1})(U^{\dag}\otimes U^{\dag})^j.
\end{equation}
When  $\ket\phi$ is the eigenstate of $O$ corresponding  to the maximum eigenvalue, the expectation value $\langle \phi |O | \phi\rangle$ attain the maximum one.
So the question is reduced to how to calculate the maximum eigenvalue of $O$.

Note that the  eigenstates of $U\otimes U$ are  states of the form $\ket{w_kw_l}$ for $k,l=0,1,2,\cdots, n-1$, and all  eigenvalues are degenerate. There has a spectral decomposition for $ U\otimes U$,
\begin{equation}
  U\otimes U= \sum_\lambda \lambda P_\lambda,
\end{equation}
where $P_\lambda$ is the projector onto the  eigenspace of $U\otimes U$ with eigenvalue $\lambda$, and $P_\lambda$ satisfy properties $\sum_\lambda P_\lambda=Id$ and $P_\lambda P_{\lambda'} =\delta_{\lambda\lambda'}P_\lambda$.
Thus the operator $O$ can be simplified as follows,
\begin{eqnarray}
 O &=& \sum_{j=0}^{nd-1}(U\otimes U)^j(\ket{0v_1^0}\bra{0v_1^0}+\ket{0v_1^1}\bra{0v_1^1})(U^{\dag}\otimes U^{\dag})^j \nonumber\\
   &=& \sum_{j=0}^{nd-1}(\sum_\lambda \lambda P_\lambda)^j(\ket{0v_1^0}\bra{0v_1^0}+\ket{0v_1^1}\bra{0v_1^1})(\sum_{\lambda'}  \lambda'^* P_{\lambda'} ^{\dag})^j \nonumber\\
   &=& \sum_{j=0}^{nd-1} \sum_\lambda \lambda^j P_\lambda (\ket{0v_1^0}\bra{0v_1^0}+\ket{0v_1^1}\bra{0v_1^1}) \sum_{\lambda'} ( \lambda'^*)^j P_{\lambda'}\nonumber\\
   &=& nd\sum_\lambda P_\lambda (\ket{0v_1^0}\bra{0v_1^0}+\ket{0v_1^1}\bra{0v_1^1}) P_{\lambda}.
\end{eqnarray}
Therefore, in order to calculate the eigenvalues of $O$, we only need to diagonalize it within the subspaces corresponding to the eigenvalues of $U\otimes U$.
Denote by $L(\lambda)$ the subspace  spanned by  all eigenvectors  $\{u_\lambda^{\lambda_j} \}$ of $U\otimes U$  corresponding to the eigenvalue $\lambda$.
The eigenvector corresponding to the maximum eigenvalue of $O$ lies in the subspace $P_\lambda (\ket{0v_1^0}\bra{0v_1^0}+\ket{0v_1^1}\bra{0v_1^1}) P_{\lambda}$ when $L(\lambda)$ has maximum dimension.

{\bf The case I}

To be connivent, we suppose that $n>1$ is odd. And choose
\begin{eqnarray}
 U &=& | w_0 \rangle\langle w_0|+e^{-i2\pi /dn} | w_1 \rangle\langle w_1|+e^{-i4\pi /dn} | w_2 \rangle\langle w_2| \nonumber\\
    &&  + \cdots + e^{i4\pi /dn} | w_{n-2} \rangle\langle w_{n-2}|+ e^{i2\pi /dn} | w_{n-1} \rangle\langle w_{n-1}|.
\end{eqnarray}

In this case, the eigenvector corresponding to the maximum eigenvalue of $O$ lies in the subspace when $\lambda=1$.

\begin{figure}[H]
  \centering
  \includegraphics[width=0.4\textwidth]{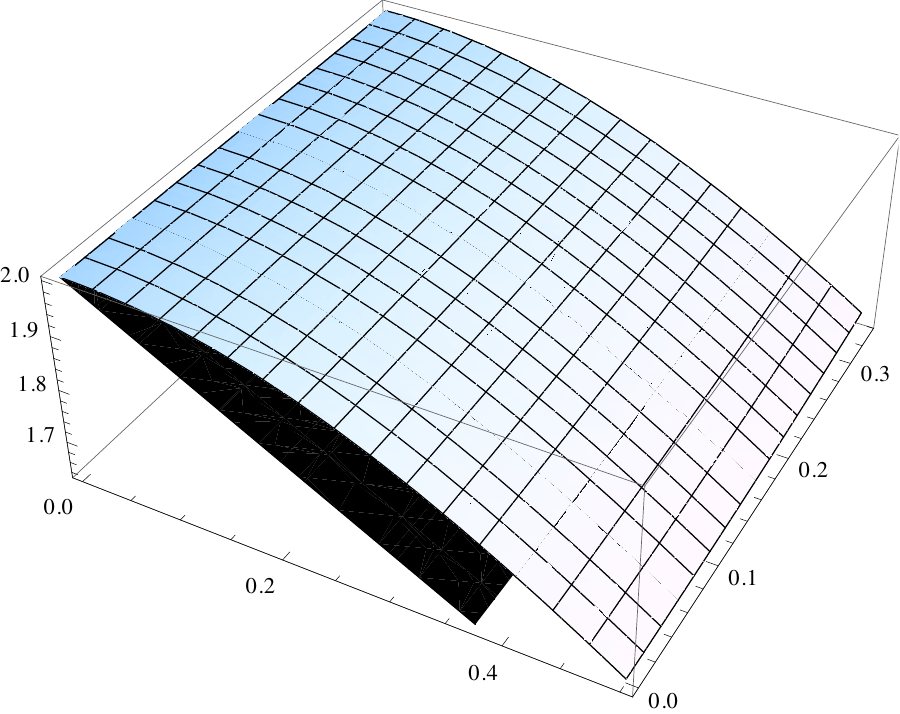}\\
  \caption{The imagine of $f(x_1,y_1)$ and $g(x_1)$ }\label{fig1}
\end{figure}

That is to say, we shall calculate the maximum eigenvalue of $P_{1}(\ket{0v_1^0}\bra{0v_1^0}+\ket{0v_1^1}\bra{0v_1^1})P_{1}$, denoting this operator by $R$.
Set
\begin{eqnarray}
  \ket{\psi_1}=P_{1}\ket{0v_1^0}, \hspace{5mm} \ket{\psi_2}=P_{1}\ket{0v_1^1}.
\end{eqnarray}
Suppose  that the eigenvectors of $R$ corresponding to  eigenvalue $\mu$ is $\sum_{j=1}^{2}x_j\ket{\psi_j}$,  then 
for any $k=1,2$,
we have $\sum_{j=1}^{2}x_j\langle \psi_k | \psi_j\rangle=\mu x_k$.
Denote the matrix $M=(\langle \psi_k | \psi_j\rangle)_{k,j=1}^2$. The eigenvalues of $M$ are exactly the ones of $R$.
With the help of Wolfram Mathematica 8.0, we  quickly obtain the matrix $M$ as the following form
\[
\left(
\begin{array}{cc}
 \frac{1}{n} & \frac{1}{n^2} (1+X) \\
 \frac{1}{n^2} (1+X) & \frac{1}{n} \\
\end{array}
\right),
\]
where $X=\cos\frac{(d-2) \pi }{2 d} \csc\frac{\pi }{d n}-\cos\frac{(d n-2) \pi }{2 d n} \csc\frac{\pi }{d n}$.
The  maximum eigenvalue of $M$ is $\lambda_{max}^M=(1+n+X)/n^2$. Thus the maximum eigenvalue $\lambda_{max}^O$ of $O$ is
\begin{eqnarray}
  \lambda_{max}^O&=&dn\lambda_{max}^M \nonumber\\
  &=& d(1+n+\cos\frac{(d-2)\pi}{2d}\csc\frac{\pi}{dn}-\cos\frac{(dn-2)\pi}{2dn}\csc\frac{\pi}{dn})/n.
\end{eqnarray}
Set $y_1=\frac{1}{n}$, $x_1=\frac{1}{d}$, then $0<y_1\leq \frac{1}{3}$, $0<x_1\leq \frac{1}{2}$. Define the functions
\begin{eqnarray}
  f(x_1,y_1) &=& 1+y_1+y_1(\sin (x_1\pi)\csc( x_1y_1\pi)-\sin( x_1y_1\pi)\csc( x_1y_1\pi)), \\
  g(x_1) &=& 2-x_1 .
\end{eqnarray}
The imagine of $f(x_1,y_1)$ and $g(x_1)$ are shown in Figure \ref{fig1} which are drawn  by
Wolfram Mathematica 8.0.
$f(x_1,y_1)$ is the curved surface and $g(x_1)$ is plane surface. Clearly, $f(x_1,y_1)>g(x_1)$ in definition  domain $0<y_1\leq \frac{1}{3}$, $0<x_1\leq \frac{1}{2}$. Equivalently, $\lambda_{max}^O>2d-1$ for any odd $n$ and any number of outcomes $d\geq 2$. Therefore, the quantum bound violates the classical bound.

{\bf The case II}

For the case $n$ is even, $n\geq 2$. We choose
\begin{eqnarray}
 U &=& | w_0 \rangle\langle w_0|+e^{-i2\pi /dn} | w_1 \rangle\langle w_1|+e^{-i4\pi /dn} | w_2 \rangle\langle w_2| + \cdots\nonumber\\
   && + e^{-i(n-2)\pi /dn} | w_{\frac{n-2}{2}} \rangle\langle w_{\frac{n-2}{2}}|+ e^{i\pi /d} | w_{\frac{n}{2}} \rangle\langle w_{\frac{n}{2}}|  +   e^{i(n-2)\pi /dn} | w_{\frac{n+2}{2}} \rangle\langle w_{\frac{n+2}{2}}|                   \nonumber\\
    &&  + \cdots + e^{i4\pi /dn} | w_{n-2} \rangle\langle w_{n-2}|+ e^{i2\pi /dn} | w_{n-1} \rangle\langle w_{n-1}|.
\end{eqnarray}

In this case, the eigenvector corresponding to the maximum eigenvalue of $O$ lies in the subspace when $\lambda=e^{i2\pi /nd}$.

We shall calculate the maximum eigenvalue of the operator
\begin{equation}\label{R2even}
  R = P_{e^{i2\pi /nd}}\ket{0v_1^0}\bra{0v_1^0}P_{e^{i2\pi /nd}}+P_{e^{i2\pi /nd}}\ket{0v_1^1}\bra{0v_1^1})P_{e^{i2\pi /nd}},
\end{equation}
Denote
\begin{eqnarray}
  \ket{\psi_1}=P_{e^{i2\pi /nd}}\ket{0v_1^0}, \hspace{5mm}
  \ket{\psi_2}=P_{e^{i2\pi /nd}}\ket{0v_1^1}, \hspace{5mm}
  M=(\langle \psi_k | \psi_j\rangle)_{k,j=1}^2,
\end{eqnarray}
 then the eigenvalues of $M$ are exactly the ones of $R$.
With the help of Wolfram Mathematica 8.0, we get the matrix $M$ has the following form
\[
\left(
\begin{array}{cc}
 \frac{1}{n} & \frac{1}{n^2} (1+X+e^{i\pi /d}) \\
 \frac{1}{n^2} (1+X+e^{-i\pi /d}) & \frac{1}{n} \\
\end{array}
\right),
\]
where $X=\cos\frac{(dn+2-2n) \pi }{2dn} \csc\frac{\pi }{d n}-\cos\frac{(d n-2) \pi }{2 d n} \csc\frac{\pi }{d n}$.
The  maximum  eigenvalue of $M$ is $\lambda_{max}^M=(n+((1+X+\cos\frac{\pi}{d})^2+(\sin\frac{\pi}{d})^2)^{1/2})/n^2$. Thus the maximum  eigenvalue $\lambda_{max}^O$ of $O$ is
\begin{eqnarray}\label{}
  \lambda_{max}^O&=&dn\lambda_{max}^M \nonumber\\
  &=& d(n+((1+X+\cos\frac{\pi}{d})^2+(\sin\frac{\pi}{d})^2)^{1/2})/n.
\end{eqnarray}

Set $y_2=\frac{1}{n}$, $x_2=\frac{1}{d}$, then $0<y_2\leq \frac{1}{2}$, $0<x_2\leq \frac{1}{2}$. Define the functions
\begin{eqnarray}
 f(x_2,y_2) & = &1+y_2((1+\widetilde{X}+\cos(x_2 \pi))^2+(\sin x_2 \pi)^2)^{\frac{1}{2}}, \\
  g(x_2) &=& 2-x_2 ,
\end{eqnarray}
where $\widetilde{X}=\sin (x_2\pi-x_2 y_2 \pi)\csc( x_2y_2\pi)-\sin( x_2y_2\pi)\csc( x_2y_2\pi)$.

\begin{figure}[H]
  \centering
  \includegraphics[width=0.4\textwidth]{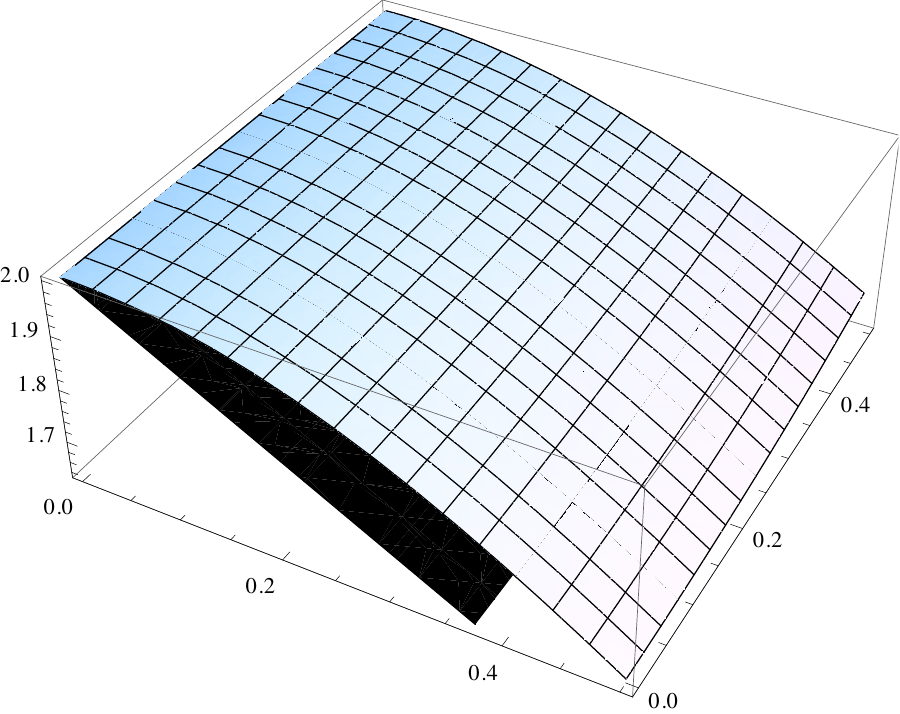}\\
  \caption{The imagine of $f(x_2,y_2)$ and $g(x_2)$ }\label{fig2}
\end{figure}

From Wolfram Mathematica 8.0, one obtain the imagine of $f(x_2,y_2)$ and $g(x_2)$ in Figure \ref{fig2}.

$f(x_2,y_2)$ is  the curved surface and $g(x_2)$ is plane surface. Clearly, in definition domain $0<y_2\leq \frac{1}{2}$, $0<x_2\leq \frac{1}{2}$, we have $f(x_2,y_2)>g(x_2)$. Equivalently, $\lambda_{max}^O>2d-1$ for any odd $n$ and any outcomes $d\geq 2$. Therefore, the quantum bound violates the classical bound.

No matter $n$ is odd or even,  the above results can be explained as a nonlocal game.
\begin{figure}[h!]
  \centering
  \includegraphics[width=0.25\textwidth]{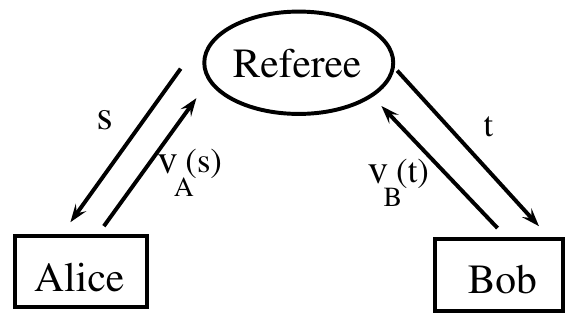}\\
  \caption{The structure of  nonlocal game}\label{nonlocal}
\end{figure}
We have that Alice and Bob each receive a bit $s$ and $t$ respectively from a referee, with each bit equally likely to be $0,\ 1,\ \cdots$, or $d-1$. After Alice and Bob perform measurements on their own part respectively,  they  send measurement results $v_A(s)$ and $v_B(t)$  back to the referee, $v_A(s)$ and $v_B(t)$ take values in the set $\{0,1,\cdots,n-1  \}$. The structure of  the nonlocal game are shown in Figure \ref{nonlocal}. The winning conditions are listed in  Table \ref{Tab1}.
\begin{table}[H]
\centering
\footnotesize{
\begin{tabular}{|c|c|c|c|c|c|c|c|c|c|}
  \hline
  s,t & 00&11& $\cdots$ & (d-1)(d-1)&01 & 12 & $\cdots$ & (d-2)(d-1)& (d-1)0 \\
  \hline
   & 01 & 01& $\cdots$ & 01 & 01 & 01& $\cdots$ & 01& 02 \\
 Alice, & 12 & 12& $\cdots$ & 12& 12 & 12& $\cdots$ & 12& 13 \\
   Bob & $\cdots$  & $\cdots$ & $\cdots$ & $\cdots$ & $\cdots$  & $\cdots$ & $\cdots$ & $\cdots$ &$\cdots$  \\
  &  (d-1)0&  (d-1)0 &$\cdots$&  (d-1)0 &  (d-1)0&  (d-1)0 &$\cdots$&  (d-1)0& (d-1)1\\
  \hline
\end{tabular}
 \caption{Winning conditions for the nonlocal game($\C^n\otimes \C^n$)}\label{Tab1}}
\end{table}

Specifically, the values of $(s,t)$ are listed on the first row, and  the corresponding bit values $v_A(s)$ and $v_B(t)$ respectively sent by Alice and Bob are listed on the second row. This game is won if the bit values $v_B(t)-v_A(s)=2\ {\text mod} \ d$ when $(s,t)=(d-1,0)$ and if the bit values $v_B(t)-v_A(s)=1\ {\text mod} \ d$ for all other allowed choice of $(s,t)$.

The maximum classical probability of winning this game can be achieved if Alice  always returning the bit value $0$ and Bob  always returning the bit value $1$. So the  maximum classical probability is $\frac{2d-1}{2d}$.

In the quantum strategy, Alice and Bob share the state $\ket\phi$ which is the eigenstate of $O$
corresponding to its maximum eigenvalue $\lambda_{max}^O$.
 If they receive  values $s$ and $t$ from the referee  respectively, Alice measure $a_s$, Bob measure $a_t$, and then they send the  measurement results to referee.
 The probability of winning this game  is then $\frac{\lambda_{max}^O}{2d}$. From Figure \ref{fig1} and Figure \ref{fig2}, we know that the value of quantum bound is larger than the value of classical bound.

Furthermore, we note that, no matter $n$ is odd or even, the classical bound of Bell inequality Eq.(\ref{Belleq O2}) is $2d-1$ which is independent of the choice of $n$. And the quantum bound is decided by both the values of $n$ and $d$.

For $n$ is odd, fix a $x_1$, we compute the partial derivative of function $f(x_1,y_1)$ with respect to $y_1$:
\begin{equation}
 \frac{\partial f(x_1,y_1) }{\partial y_1}=\csc(\pi x_1 y_1) \sin(\pi x_1) - \pi x_1 y_1 \cot(\pi x_1 y_1) \csc(\pi x_1 y_1)\sin(\pi x_1).
\end{equation}
In the domain  $0<y_1\leq \frac{1}{3}$, $0<x_1\leq \frac{1}{2}$, the trigonometric functions $\csc(\pi x_1 y_1)$, $\sin (\pi x_1)$ and $\cot(\pi x_1 y_1)$ are all exceed $0$, so the derivative function $\frac{\partial f(x_1,y_1) }{\partial y_1}$ always exceed $0$ in the definition domain. Thus the continuous function  $f(x_1,y_1)$ is a monotonic increasing function for any fixed $x_1$. That is to say, the quantum bound $\lambda_{max}^O$ is a monotonic decreasing  function with respect to $n$ for a fixed $d$, the number of measurements.

\begin{figure}[H]
  \centering
  \includegraphics[width=0.4\textwidth]{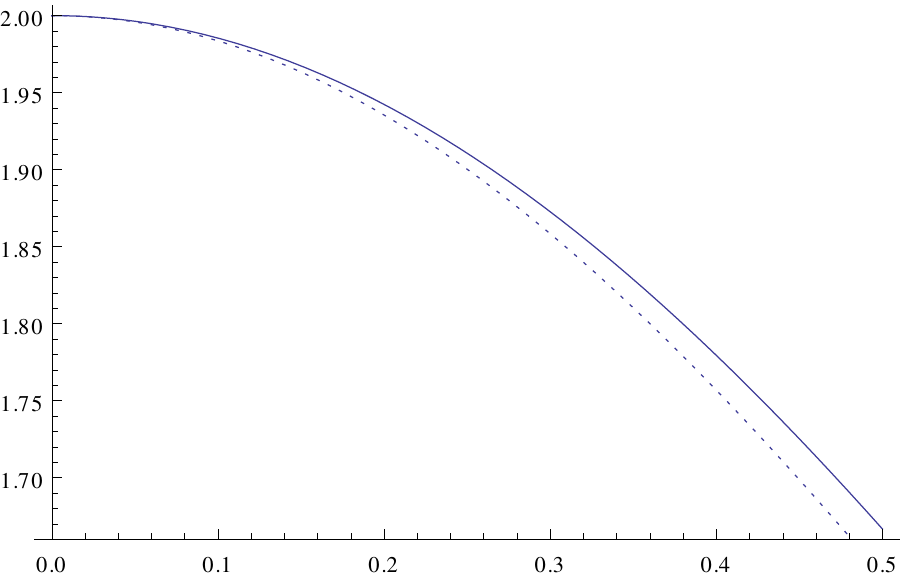}\\
  \caption{The imagine of $f(x_1,0)$ and $f(x_1,\frac{1}{3})$ }\label{fig3}
\end{figure}

Even so, when $d$ is large enough, the quantum bound is independent of the choice of $n$.
If each measurement has $3$ outcomes, i.e. $y_1=\frac{1}{3}$,
\begin{eqnarray}
  f(x_1,\frac{1}{3}) &=& 1+\frac{1}{3}+\frac{1}{3}(\sin (x_1\pi)\csc \frac{x_1\pi}{3}-\sin \frac{x_1\pi}{3}\csc \frac{x_1\pi}{3}) \nonumber\\
   &=& \frac{2}{3} (2+\cos\frac{2 \pi  x}{3}).
\end{eqnarray}
If $n$ is large enough, i.e. $y_1\rightarrow 0$, we evaluate the limit value
\begin{eqnarray}
  f(x_1,0) &:=& \lim_{y_1\rightarrow 0}f(x_1,y_1) \nonumber\\
   &=& 1+\frac{1}{\pi  x}\sin(\pi  x).
\end{eqnarray}
We draw the graphics of functions $f(x_1,\frac{1}{3})$ and $f(x_1,0)$ in Figure \ref{fig3},
function $f(x_1,\frac{1}{3})$ is solid, and function $f(x_1,0)$ is dotted.  From this figure, we see that when $x_1$ is near to $0$, $f(x_1,0)$ approach $f(x_1,\frac{1}{3})$.

When $n$ is even, we have similar analysis. That is to say, when the number of measurements is  large enough, the violation of Bell inequality is determined by the number of measurements $d$ and independent of $n$, the number of outcomes.

\section{Four partite system}\label{sec3}

For a four partite  system, Alice, Bob, Charlie  and Danniel share  joint state $\ket\psi$ which are elements of the Hilbert space   $\mathbb{C}^4\otimes \C^4 \otimes \C^4 \otimes \C^4$, with each $\C^4$ is spanned by the orthonormal basis $\{\ket0, \ket1, \ket2, \ket3  \}$.
Each of them can take one of two measurements, and for each measurement the possible values for the outcomes are $0$, $1$, $2$ or $3$.  Alice's observable  operators are $a_{0}$ and $a_1$,
Bob's are $b_{0}$ and $b_{1}$,  Charlie's are $c_{0}$ and $c_{1}$, and Danniel's are $d_{0}$ and $d_{1}$.
The orthonormal basis $\{\ket0, \ket1, \ket2, \ket3  \}$ correspond to the observable  operators $a_0$, $b_0$, $c_0$ and $d_0$, $a_0=|1\rangle\langle 1| + 2|2\rangle\langle 2|+3|3\rangle\langle 3|$, similarly for $b_0$, $c_0$ and $d_0$. Next we define the second basis.

Since the $4\times 4$ translation operator $T$ is an orthogonal matrix under any orthonormal basis, we have
\begin{equation}\label{T4}
  T= | w_0 \rangle\langle w_0| +e^{-i\pi /2} | w_1  \rangle\langle w_1 |+e^{i\pi} | w_2  \rangle\langle w_2 |+ e^{i\pi /2}| w_3  \rangle\langle w_3 |,
\end{equation}
where
\begin{equation}\label{w_j4}
|w_j \rangle=\frac{1}{2}\sum_{k=0}^3 e^{i\pi jk /2}\ket j, \hspace{5mm} \langle w_j|w_k\rangle=\delta_{jk}
\end{equation}
for any $j,k=0,1,2,3$.
Choose operator
\begin{equation}\label{U}
  U= | w_0 \rangle\langle w_0| +e^{-i\pi /4} | w_1  \rangle\langle w_1 |+e^{i\pi /2} | w_2  \rangle\langle w_2 |+ e^{i\pi /4}| w_3  \rangle\langle w_3 |,
\end{equation}
then the space $\{j\rightarrow U^j \mid j=0,1,\cdots, 7 \}$ is a representation of the cyclic group $\Z_8$, the group of integers $\Z$ modulo $8$.
We can define the second basis $\{ v_j=U^j \ket j \mid j=0,1,2,3\}$, and this basis corresponds to the observable operators $a_1$, $b_1$, $c_1$ and $d_1$, $a_1=|v_1\rangle\langle v_1| + 2|v_2\rangle\langle v_2|+3|v_3\rangle\langle v_3|$, similarly for $b_1$, $c_1$ and $d_1$.
Denote $V=U\otimes U\otimes U\otimes U$, then $\{j\rightarrow V^j  \mid j=0,1,\cdots, 7 \}$ is also a representation of the cyclic group $\Z_8$.
We choose the following six states:
\begin{equation}\label{6orbits}
\ket{000v_1}, \hspace{5mm} \ket{0v_0v_00}, \hspace{5mm} \ket{0v_00v_0}, \hspace{5mm} \ket{v_0v_303}, \hspace{5mm} \ket{0v_1v_11}, \hspace{5mm} \ket{0v_220},
\end{equation}
where state $\ket{000v_1}$ means $\ket 0 \otimes \ket 0 \otimes\ket 0 \otimes\ket {v_1}$. Each of these six states in (\ref{6orbits}) has an orbit with $8$ elements under the action of $\{ V^j | j=0,1,\cdots, 7 \}$, and the six orbits are distinct.
The set of all states in the six orbits leads to a  violation of Bell inequality. The specific analysis is given as follows.

For each state in the six orbits, it corresponds to a particular choice of measurements  and  the corresponding measurement results. For example, for the state  $\ket{000v_1}$, it corresponds to the fact that Alice measuring $a_0$ and obtaining $0$,  Bob measuring $b_0$ and obtaining $0$,  Charlie  measuring $c_0$ and obtaining $0$,  Danniel  measuring $d_1$ and obtaining $1$.

From local realistic theory,
the sum of  these $48$ events give a  Bell inequality, it reads
\begin{equation}
\footnotesize{
 \begin{split}
&\sum_{j=0}^3 P(a_0=b_0=c_0=j,d_1=j\oplus1)+\sum_{j=0}^3 P(a_1=b_1=c_1=j,d_0=j\oplus2)\\
& +\sum_{j=0}^3 P(a_0=b_1=c_1=d_0=j) +\sum_{j=0}^3 P(a_1=d_1=j,b_0=c_0=j\oplus1)\\
& +\sum_{j=0}^3 P(a_0=b_1=c_0=d_1=j) +\sum_{j=0}^3 P(a_1=c_1=j,b_0=d_0=j\oplus1) \\
& +\sum_{j=0}^3 P(a_1=c_0=j,b_1=d_0=j\oplus3) +\sum_{j=0}^3 P(a_0=j,b_0=c_1=j\oplus3,d_1=j\oplus2)\\
& +\sum_{j=0}^3 P(a_0=d_0=j,b_1=c_1=j\oplus1) +\sum_{j=0}^3 P(a_1=d_1=j,b_0=c_0=j\oplus1) \\
& +\sum_{j=0}^3 P(a_0=d_0=j,b_1=c_0=j\oplus2) +\sum_{j=0}^3 P(a_1=d_1=j,b_0=j\oplus3,c_1=j\oplus2)\leq 2,
\end{split}}
\end{equation}
where $\oplus$ means the addition modulo $4$,
$P(a_0=b_0=c_0=j,d_1=j\oplus1)$ means the probability of the event when Alice, Bob and Charlie take measure $a_0$, $b_0$ and $c_0$ respectively and obtain the same value $j$, Danniel  take measure $d_1$ and obtain  value $j\oplus1$.
But in  quantum  bound  can attain the value $2.021$.

In order to maximize the sum of probabilities corresponding to these $48$ states in  quantum mechanics, we need to find a  state $\ket\phi$, such that the expectation value $\langle \phi |O | \phi\rangle$ is maximum, where
\begin{equation}\label{expect O4}
  O= \sum_{j=0}^7V^jL(V^{\dag})^j,
\end{equation}
and
\begin{equation}\label{expect L4}
\begin{split}
&  L=\ket{000v_1}\bra{000v_1}+\ket{0v_0v_00}\bra{0v_0v_00}+\ket{0v_00v_0}\bra{0v_00v_0}+\\
&\qquad  \ket{v_0v_303}\bra{v_0v_303}+\ket{0v_1v_11}\bra{0v_1v_11}+\ket{0v_220}\bra{0v_220}.
\end{split}
\end{equation}
The maximum value of $\langle \phi |O | \phi\rangle$ occurs when $\ket\phi$ is the eigenstate of $O$ corresponding  to its maximum eigenvalue. So the question is reduced to how to calculate the maximum eigenvalue of $O$.

Note that the  eigenstates of $V$ are  states of the form $\ket{w_kw_lw_mw_n}$ for $k,\ l,\ m,\ n=0,\ 1,\ 2,\ 3$, and the eigenvalues are $\pm1$, $e^{\pm i\pi/4}$, $e^{\pm i\pi/2}$ and $e^{\pm 3i\pi/4}$. The spectral decomposition
$V= \sum_\lambda \lambda P_\lambda$,
 leads to a simplification of $O$,
$O = 8\sum_\lambda P_\lambda L P_{\lambda}.$
The eigenvector corresponding to the maximum eigenvalue of $O$ lies in the subspace when $V$ has eigenvalue $e^{i\pi/2}$.
Set
\begin{eqnarray*}
  \ket{\psi_1}=P_{e^{i\pi/2}}\ket{000v_1}, \hspace{5mm} \ket{\psi_2}=P_{e^{i\pi/2}}\ket{0v_0v_00}, \hspace{5mm} \ket{\psi_3}=P_{e^{i\pi/2}}\ket{0v_00v_0}, \nonumber\\ \ket{\psi_4}=P_{e^{i\pi/2}}\ket{v_0v_303}, \hspace{5mm} \ket{\psi_5}=P_{e^{i\pi/2}}\ket{0v_1v_11}, \hspace{5mm}
  \ket{\psi_6}=P_{e^{i\pi/2}}\ket{0v_220},
\end{eqnarray*}
and $R=P_{e^{i\pi/2}}LP_{e^{i\pi/2}}$,
then the operator can be written as,
\begin{equation}\label{R4}
  R = \sum_{j=1}^6 \ket{\psi_j}\bra{\psi_j}.
\end{equation}

Note that the eigenvectors of $R$ can be expressed as $\sum_{j=1}^{6}x_j\ket{\psi_j}$, then there exists  eigenvalue $\mu$ such that $\sum_{k=1}^6 \ket{\psi_k}\bra{\psi_k}\sum_{j=1}^{6}x_j\ket{\psi_j}=\mu\sum_{j=1}^{6}x_j\ket{\psi_j}$. Rewrite the  eigenvalue equation, it becomes
\begin{equation}
  \sum_{j=1}^{6}x_j\langle \psi_k | \psi_j\rangle\sum_{k=1}^{6}\ket{\psi_k}=\mu\sum_{k=1}^{6}x_k\ket{\psi_k}.
\end{equation}
Equivalently, for any $k=1,2,\cdots,6$, we have $\sum_{j=1}^{6}x_j\langle \psi_k | \psi_j\rangle=\mu x_k$.
Denote the matrix $M=(\langle \psi_k | \psi_j\rangle)_{k,j=1}^6$, then the eigenvalues of $M$ are exactly the ones of $R$.
With the help of Wolfram Mathematica 8.0, we quickly obtain the matrix $256M$ as the following form
\[
\tiny{
\left(
\begin{array}{cccccc}
\medskip
 44 & -2-\sqrt{2}+ i\sqrt{2} & -2-\sqrt{2}+ i\sqrt{2} & -\sqrt{2}-i(2+\sqrt{2}) & 4-3\sqrt{2}+i(\sqrt{2}-2) & 4-4i \\
 \medskip
 -2-\sqrt{2}- i\sqrt{2} & 44 & 8+4\sqrt{2} & i 4\sqrt{2} & 8i & -\sqrt{2}+ i(2-\sqrt{2}) \\
 \medskip
 -2-\sqrt{2}- i \sqrt{2}& 8+4\sqrt{2} & 44 & 0 & 0 & -\sqrt{2}+ i(2-\sqrt{2}) \\
 \medskip
 -\sqrt{2}+i(2+\sqrt{2})& - i 4\sqrt{2} & 0 & 44 & 0 & -2+\sqrt{2}- i\sqrt{2} \\
 \medskip
 4-3\sqrt{2}- i(\sqrt{2}-2) & -8 i & 0 & 0 & 44 & 2+\sqrt{2}+ i(4+3\sqrt{2}) \\
 4+4 i & -\sqrt{2}- i(2-\sqrt{2}) & -\sqrt{2}- i(2-\sqrt{2}) & -2+\sqrt{2}+ i\sqrt{2} & 2+\sqrt{2}- i(4+3 \sqrt{2}) & 44 \\
\end{array}
\right),}
\]
and obtain the numerical approximate value of the maximum eigenvalue of $M$ is $64.667/256$. Thus the maximum eigenvalue of $O$ is $8\times\frac{64.667}{256}\approx2.021>2$. So we get a violation.

The above results can be explained as a nonlocal game.  The bits $s$, $t$, $u$  and $v$ are sent to Alice, Bob, Charlie and  Danniel respectively from the same referee, take value $0$ or $1$ with equally likely possibility. Alice take measure $a_s$ on her part, Bob take measure $b_t$ on his part, Charlie take measure $c_u$ on her part,  Danniel take measure $d_v$ on his part. Then they each send a bit value $v_A(s)$, $v_B(t)$, $v_C(u)$ and $v_D(v)$  back to the referee, the bit values take values in the set $\{0,\ 1,\ 2,\ 3\}$. The winning conditions are listed in  Table \ref{Tab2}.
\begin{table}[h!]
\begin{minipage}{.45\linewidth}
\centering
\small{
\begin{tabular}{|c|c|}
  \hline
  s,t,u,v  &  Alice, Bob, Charlie,  Danniel \\
  \hline
  0001 & 0001, 1112, 2223, 3330 \\
  1110 & 0002, 1113, 2220, 3331 \\
  0110 & 0000, 1111, 2222, 3333 \\
  1001 & 0110, 1221, 2332, 3003 \\
  0101 & 0000, 1111, 2222, 3333  \\
  1010 & 0101, 1212, 2323, 3030  \\
  \hline
\end{tabular}}
\end{minipage}
\begin{minipage}{.45\linewidth}
\centering
\small{
\begin{tabular}{|c|c|}
  \hline
  s,t,u,v  &  Alice, Bob, Charlie, Danniel \\
  \hline
  1100 & 0303, 1010, 2121, 3232 \\
  0011 & 0332, 1003, 2110, 3221 \\
  0110 & 0111, 1222, 2333, 3000 \\
  1001 & 0221, 1332, 2003, 3110 \\
  0100 & 0220, 1331, 2002, 3113  \\
  1011 & 0320, 1031, 2102, 3213  \\
  \hline
\end{tabular}}
\end{minipage}
 \caption{Winning conditions for the nonlocal game(4parties)}\label{Tab2}
\end{table}

Specifically, the values of $(s,t,u,v)$ are listed on the left, and  the corresponding outcomes $v_A(s)$, $v_B(t)$, $v_C(u)$ and $v_D(v)$ sent by  Alice, Bob, Charlie and Danniel are listed on the right.
The maximum classical probability of winning this game can be achieved when the outcomes
$v_A(s)$, $v_B(t)$, $v_C(u)$ and $v_D(v)$ take same value,
$(s,t,u,v)$ take value $(0110)$ or $(0110)$. So the  maximum classical probability is $\frac{2}{12}\approx 0.1667$.

In the quantum strategy, Alice, Bob, Charlie and Danniel share the state $\phi$ which is the eigenstate of $O$
corresponding to its maximum eigenvalue $2.021$.
 If they receive  values $s$, $t$, $u$  and $v$ from the referee  respectively, Alice measure $a_s$, Bob measure $a_t$, Charlie measure $a_u$ and Danniel measure $a_v$,  then they send the  measurement results to referee.
 With this strategy, the probability of winning this game  is $0.1684$.

\section{Three partite system}\label{sec4}
Alice, Bob and Charlie make measurements, each party make one of $2$ possible measurements, and each measurement has $4$ outcomes.
As above, the orthonormal basis $\{\ket0, \ket1, \ket2, \ket3  \}$ correspond to the observables $a_0$, $b_0$ and $c_0$, $a_0=|1\rangle\langle 1| + 2|2\rangle\langle 2|+3|3\rangle\langle 3|$, similarly for $b_0$ and $c_0$. The orthonormal basis $\{\ket{v_0}, \ket{v_1}, \ket{v_2}, \ket{v_3} \}$ correspond to the observables $a_1$, $b_1$ and $c_1$, $a_1=|v_1\rangle\langle v_1| + 2|v_2\rangle\langle v_2|+3|v_3\rangle\langle v_3|$, similarly for $b_1$ and $c_1$.

We restrict the six orbits in Eq.(\ref{6orbits}) to three partite system by forgetting the last party, and guarantee the bound of probabilities from local realistic
theory invariant. We get  $4$ orbits in $\C^4\otimes\C^4\otimes\C^4$, which will get a violation of Bell inequality. The  representative elements of the $4$ orbits are:
\begin{equation}\label{4orbits}
 \ket{0v_0v_0}, \hspace{5mm}  \ket{v_0v_30}, \hspace{5mm} \ket{0v_1v_1}, \hspace{5mm} \ket{0v_22}.
\end{equation}
 Each of the four states in Eq.(\ref{4orbits}) has an
orbit with $8$ elements under the action of $\{U^j\otimes U^j\otimes U^j| j=0,1, \cdots, 7  \}$, and the four orbits are distinct. From local realistic theory, the sum of these $32$ states give a Bell inequality. It reads
\begin{equation}
\footnotesize{
 \begin{split}
&\sum_{j=0}^3 P(a_0=b_1=c_1=j)+\sum_{j=0}^3 P(a_1=j,b_0=c_0=j\oplus1)\\
& +\sum_{j=0}^3 P(a_1=c_0=j,b_1=j\oplus3) +\sum_{j=0}^3 P(a_0=j,b_0=c_1=j\oplus3)\\
& +\sum_{j=0}^3 P(a_0=j,b_1=c_1=j\oplus1) +\sum_{j=0}^3 P(a_1=j,b_0=c_0=j\oplus2) \\
& +\sum_{j=0}^3 P(a_0=j,b_1=c_0=j\oplus2) +\sum_{j=0}^3 P(a_1=j,b_0=j\oplus3,c_1=j\oplus2)\leq 2,
\end{split}}
\end{equation}
where $\oplus$ means the addition modulo $4$.
But the  quantum  bound of the sum of these probabilities can attain the value $2.075$.

Similarly,  we need to find a special state $\ket\phi$, such that the expectation value $\langle \phi |O | \phi\rangle$ is maximum, where
\begin{equation}\label{expect O3}
  O= \sum_{j=0}^7(U\otimes U\otimes U)^jL(U^{\dag}\otimes U^{\dag}\otimes U^{\dag})^j,
\end{equation}
and
\begin{equation}\label{expect L3}
L=\ket{0v_0v_0}\bra{0v_0v_0}+ \ket{v_0v_30}\bra{v_0v_3}+\ket{0v_1v_1}\bra{0v_1v_1}+\ket{0v_22}\bra{0v_22}.
\end{equation}

Note that the  eigenstates of $U\otimes U\otimes U$ are  states of the form $\ket{w_kw_mw_n}$ for $k,\ m,\ n=0,\ 1,\ 2,\ 3$, and the eigenvalues are $\pm1$, $e^{\pm i\pi/4}$, $e^{\pm i\pi/2}$ and $e^{\pm 3i\pi/4}$. For the spectral decomposition of $U$,
$ U\otimes U\otimes U= \sum_\lambda \lambda P_\lambda$,
the operator $O$ can be simplified as
\begin{equation}
  O = 8\sum_\lambda P_\lambda L P_{\lambda}.
\end{equation}
The eigenvector corresponding to the maximum eigenvalue of $O$ lies in the subspace when $U\otimes U\otimes U$ has eigenvalue $e^{i\pi/2}$.
Set
\begin{eqnarray}
  \ket{\psi_1}=P_{e^{i\pi/2}}\ket{0v_0v_0}, \hspace{5mm} \ket{\psi_2}=P_{e^{i\pi/2}}\ket{v_0v_30}, \nonumber \\ \ket{\psi_3}=P_{e^{i\pi/2}}\ket{0v_1v_1}, \hspace{5mm} \ket{\psi_4}=P_{e^{i\pi/2}}\ket{0v_22}.
\end{eqnarray}
Denote the operator $P_{e^{i\pi/2}}LP_{e^{i\pi/2}}$ by  $R$, then $R$ can be write as
$R = \sum_{j=1}^4 \ket{\psi_j}\bra{\psi_j}.$
Suppose  that the eigenvectors of $R$ corresponding to  eigenvalue $\mu$ is $\sum_{j=1}^{4}x_j\ket{\psi_j}$, then the  eigenvalue equation can be rewritten  as
\begin{equation}\label{}
  \sum_{j=1}^{4}x_j\langle \psi_k | \psi_j\rangle\sum_{k=1}^{4}\ket{\psi_k}=\mu\sum_{k=1}^{4}x_k\ket{\psi_k}.
\end{equation}
Denote the matrix $M=(\langle \psi_k | \psi_j\rangle)_{k,j=1}^4$.  The eigenvalues of $M$ are exactly the ones of $R$.
With the help of Mathematics, we  obtain the matrix $M$ as the following form
\[
\small{
\frac{1}{64}\left(
\begin{array}{cccc}
 12 & i \sqrt{2} & 2 i & 1+i \\
 -i \sqrt{2} & 12 & -1-i & \sqrt{2} \\
 -2 i & -1+i & 12 & 2 i+i \sqrt{2} \\
 1-i & \sqrt{2} & -2 i-i \sqrt{2} & 12 \\
\end{array}
\right),}
\]
and obtain the numerical approximate value of the maximum eigenvalue of $M$ is $16.597/64$. Thus the maximum eigenvalue of $O$ is $8\times\frac{16.597}{64}\approx2.075>2$.

This results can also be phrased as a nonlocal game.  Alice, Bob and Charlie each receive a bit $s$, $t$ and $u$  respectively from same referee.  $s$, $t$ and $u$ take value  $0$ or $1$ with equally likely possibility. Then Alice make measurement $a_s$, Bob make measurement $b_t$ and Charlie make measurement $c_u$. They return the measurement results $v_A(s)$, $v_B(t)$ and $v_C(u)$ back to the referee respectively, the outcomes take values in the set $\{0,\ 1,\ 2,\ 3\}$. The winning conditions are listed in  Table \ref{Tab3}.
\begin{table}[H]
\begin{minipage}{.45\linewidth}
\centering
\small{
\begin{tabular}{|c|c|}
  \hline
  s,t,u  & Alice, Bob, Charlie  \\
  \hline
  011 & 000, 111, 222, 333 \\
  100 & 011, 122, 233, 300 \\
  110 & 030, 101, 212, 323 \\
  001 & 033, 100, 211, 322 \\
  \hline
\end{tabular}}
\end{minipage}
\begin{minipage}{.45\linewidth}
\centering
\small{
\begin{tabular}{|c|c|}
  \hline
  s,t,u  &  Alice, Bob, Charlie \\
  \hline
  011 & 011, 122, 233, 300 \\
  100 & 022, 133, 200, 311 \\
  010 & 022, 133, 200, 311  \\
  101 & 032, 103, 210, 321  \\
  \hline
\end{tabular}}
\end{minipage}
 \caption{Winning conditions for the nonlocal game(3parties)}\label{Tab3}
\end{table}

Specifically, the values of $(s,t,u)$ are listed on the left, and  the corresponding measurement results $v_A(s)$, $v_B(t)$ and $v_C(u)$ sent by Alice, Bob and Charlie are listed on the right.
Note that, the states listed in (\ref{4orbits}) is obtained by the states in (\ref{6orbits}) by forgetting the last bit and an additional condition. The Table \ref{Tab3} is obtained by deleting the first two rows and the last two rows in the left table of Table \ref{Tab2}, and forgetting the last bit values.
The maximum classical probability of winning this game  is $\frac{2}{8}\approx 0.25$.

In the quantum strategy, Alice, Bob and Charlie share the state $\phi$ which is the eigenstate of $O$ corresponding to its maximum eigenvalue $2.075$.
If they receive  values $s$, $t$ and $u$   from the referee  respectively, then Alice measure $a_s$, Bob measure $a_t$,  Charlie measure $a_u$,  and they send the  measurement results to referee.
 With this strategy, the probability of winning this game  is $0.2594$.

\subsection*{Acknowledgement}
This work is supported by NSFC 11571119 and NSFC 11475178.

\end{document}